\documentclass[11pt]{amsart}
\usepackage{amsmath,amsthm,amssymb}
\usepackage{url}
\usepackage{bm}

\usepackage{enumerate}

\title{Approximation Algorithms for Flexible Graph Connectivity}
\thanks{
A preliminary version of this paper
appeared in the Proceedings of the
{41st IARCS Annual Conference on Foundations of Software Technology and Theoretical Computer Science (FSTTCS~2021)},
{December 15--17, 2021},
Ed.~{M.~Boja\'{n}czyk and C.~Chekuri}
(LIPIcs, Volume 213, Article No.~9, pp.~9:1--9:14).
}

\date{}

\author{Sylvia Boyd}
\address{School of Electrical Engineering \& Computer Science, University of Ottawa, Canada}
\email{sboyd@uottawa.ca} 
\urladdr{https://www.site.uottawa.ca/~sylvia}
\author{Joseph Cheriyan}
\address{Department of Combinatorics and Optimization, University of Waterloo, Canada}
\email{jcheriyan@uwaterloo.ca}
\urladdr{http://www.math.uwaterloo.ca/~jcheriyan}
\thanks{The second author is supported in part by NSERC, RGPIN-2019-04197.}
\author{Arash Haddadan}
\address{Warner Music Group, New York, NY, USA}
\email{arash.haddadan@gmail.com}
\thanks{This work of the third author was mostly done as a postdoctoral researcher at the Biocomplexity Institute and Initiative at the University of Virginia, Charlottesville, and supported by the NSF Expeditions in Computing Grant with award number CCF-1918656.}
\author{Sharat Ibrahimpur}
\address{Department of Combinatorics and Optimization, University of Waterloo, Canada}
\email{sharat.ibrahimpur@uwaterloo.ca}
\urladdr{http://www.math.uwaterloo.ca/~s26ibrah \quad {https://orcid.org/0000-0002-1575-9648}}
\thanks{The fourth author is supported in part by NSERC grant 327620-09.}



\newtheorem{theorem}{Theorem}[section]
\newtheorem{proposition}[theorem]{Proposition}
\newtheorem{lemma}[theorem]{Lemma}

\newtheorem{claim}[theorem]{Claim}
\newtheorem{definition}{Definition}[section]

\usepackage[mathscr]{euscript}
\usepackage{tikz}
\newcommand\IGNORE[1]{}

\newcommand{\Rp}{\ensuremath{\mathbb R_{\geq 0}}}
\newcommand{\Zint}{\ensuremath{\mathbb Z}}
\newcommand{\Zp}{\ensuremath{\mathbb Z_{\geq 0}}}

\newcommand{\del}{\ensuremath{\delta}}
\newcommand{\din}{\ensuremath{\del^{\mathrm{in}}}}

\newcommand{\opt}{\textsc{OPT}}
\newcommand{\safe}{\mathscr{S}}
\newcommand{\unsafe}{\mathscr{U}}
\newcommand{\C}{\mathcal{C}}
\newcommand{\E}{\mathcal{E}}
\newcommand{\Rc}{\mathcal{R}}
\newcommand{\tjn}{{W}}

\newcommand{\mst}{\mathrm{MST}}
\newcommand{\fgc}{\mathrm{FGC}}
\newcommand{\ecss}{\mathrm{ECSS}}
\newcommand{\kecss}[1][k]{#1\text{-}\ecss}
\newcommand{\twoecss}{2\text{-}\ecss}
\newcommand{\pqfgc}{(p,q)\text{-}\fgc}
\newcommand{\onefgc}{(1,1)\text{-}\fgc}
\newcommand{\ourfgc}{(k,1)\text{-}\fgc}
\newcommand{\ahmfgc}{(1,k)\text{-}\fgc}

\newcommand{\capkecss}[1][k]{\mathrm{Cap}\text{-}\kecss[#1]}

\newcommand{\unfn}{f}

\newcommand{\disjcup}{ \,\dot\cup\, }

\newcommand{\umax}[1][u]{{#1}_{\mathrm{max}}}
\newcommand{\umin}[1][u]{{#1}_{\mathrm{min}}}

\newenvironment{proofof}[1]{\begin{proof}[{Proof of #1}]}{\end{proof}}

\begin{document}

\begin{abstract}
{
We present approximation algorithms for several network design problems
in the model of Flexible Graph Connectivity
(Adjiashvili, Hommelsheim and M\"uhlenthaler, ``Flexible Graph Connectivity'',
\textit{Math.~Program.} pp.~1--33 (2021), \textit{IPCO}~2020: pp.~13--26).

Let $k\geq 1$, $p\geq 1$ and $q\geq 0$ be integers.
In an instance of the $(p,q)$-Flexible Graph Connectivity problem, denoted $\pqfgc$, 
we have an undirected connected graph $G = (V,E)$, a partition
of $E$ into a set of safe edges $\safe$ and a set of unsafe edges
$\unsafe$, and nonnegative costs $c: E\to\Rp$ on the edges.
A subset $F \subseteq E$ of edges is feasible for the $\pqfgc$ problem if
for any set $F'\subseteq\unsafe$ with $|F'|\leq q$,
the subgraph $(V, F \setminus F')$ is $p$-edge connected.
The algorithmic goal is to find a feasible solution $F$ that 
minimizes $c(F) = \sum_{e \in F} c_e$.
We present a simple $2$-approximation algorithm for the $\onefgc$ problem
via a reduction to the minimum-cost rooted $2$-arborescence problem.
This improves on the $2.527$-approximation algorithm of Adjiashvili et~al.
Our $2$-approximation algorithm for the $\onefgc$ problem extends to a
$(k+1)$-approximation algorithm for the $\ahmfgc$ problem.
We present a $4$-approximation algorithm for the $\ourfgc$ problem,
and an $O(q\log|V|)$-approximation algorithm for the $\pqfgc$ problem.
Finally, we improve on the result of Adjiashvili et~al.\ for the
unweighted~$\onefgc$ problem by presenting a $16/11$-approximation algorithm.

The $\pqfgc$ problem is related to
the well-known \textit{Capacitated $k$-Connected Subgraph} problem (denoted $\capkecss$)
that arises in the area of Capacitated Network Design.
We give a $\min(k,2 \umax)$-approximation algorithm for the
$\capkecss$ problem, where $\umax$ denotes the maximum capacity of an edge.

\bigskip
\bigskip
\noindent
Corresponding Author: Joseph Cheriyan \\
Affiliation:
{Department of Combinatorics and Optimization, University of Waterloo, Canada} \\
E-mail Address: \texttt{jcheriyan@uwaterloo.ca}
}
\end{abstract}

\maketitle

\clearpage

{
\section{Introduction} \label{sec:intro}
Network design and graph connectivity are core topics in 
Theoretical Computer Science and Operations Research.
A basic problem in network design is to find a minimum-cost sub-network $H$ 
of a given network $G$ such that $H$ satisfies some specified connectivity 
requirements. Most of these problems are NP-hard.
Several important algorithmic paradigms were developed in the context
of these topics, ranging from exact algorithms for the shortest
${s,t}$-path problem and the minimum spanning tree ($\mst$) problem
to linear programming-based approximation algorithms for the
survivable network design problem and the generalized Steiner network problem.
Network design problems are often motivated by practical
considerations such as the design of fault-tolerant supply chains,
congestion control for urban road traffic, and the modeling of
epidemics (see \cite{MW84,RGPV16,SSDC14}).

Recently, Adjiashvili, Hommelsheim and M\"uhlenthaler \cite{AHM20,AHM21}
introduced a new model called \textit{Flexible Graph Connectivity} ($\fgc$),
that is motivated by research in robust optimization.
(In this paper, the notation $\fgc$ may be used as an abbreviation for
``the $\fgc$ problem''; we use similar abbreviations for the names of
other related problems.)
In an instance of $\fgc$, we have an undirected connected graph
$G = (V,E)$, a partition of $E$ into a set of safe edges
$\safe$ and a set of unsafe edges $\unsafe$, and
nonnegative costs $c: E\to\Rp$ on the edges.
The graph $G$ may have multiedges, but no self-loops.
We use $n$ to denote the number of vertices of $G$.
The cost of an edge-set $F\subseteq{E}$ is denoted by $c(F) = \sum_{e\in{F}} c_e$.
A subset $F \subseteq E$ of edges is feasible for $\fgc$ if
for any unsafe edge $e \in F \cap \unsafe$,
the subgraph $(V,F \setminus \{e\})$ is connected.
The problem is to find a feasible edge-set $F$ of minimum cost.
The motivation for studying $\fgc$ is two-fold.
First, $\fgc$ generalizes many well-studied survivable network design problems.
Notably, the problem of finding a minimum-cost $2$-edge connected
spanning subgraph (abbreviated as $\twoecss$) corresponds to 
the special case of $\fgc$ where all edges are unsafe, and the $\mst$~problem
corresponds to the special case of $\fgc$ where all edges are safe.
Second, $\fgc$ captures a non-uniform model of survivable network design problems
where a specified subset of edges never fail, whereas
each edge can fail in the classical model of survivable network design problems.
Since $\fgc$ generalizes the minimum-cost $\twoecss$ problem, it
is APX-hard (see \cite{GGTW09}); thus, a polynomial-time approximation
scheme for $\fgc$ is ruled out unless P$=$NP.

The notion of $\pqfgc$ is an extension of the basic $\fgc$ model
where we have two additional integer parameters $p$ and $q$ satisfying
$p \geq 1$ and $q \geq 0$.
For a subgraph $H=(V,F)$ of $G$ and a vertex-set
$S\subseteq{V}$, we use $\del_{H}(S)$ to denote the set of edges
in $H$ with exactly one end-vertex in $S$.
A subset $F \subseteq E$ of edges is feasible for $\pqfgc$ if
the spanning subgraph $H=(V,F)$ is $p$-edge connected, and moreover,
the deletion of any set of at most $q$ unsafe edges of $F$ preserves $p$-edge connectivity.
In other words, each cut $\del_H(S)$, $\emptyset\neq{S}\subsetneq{V}$, of $H$ either
contains $p$ safe~edges or contains $p+q$ (safe or unsafe) edges.
The algorithmic goal is to find a feasible edge-set $F$ of minimum cost.
The $\pqfgc$ problem is a natural and fundamental question in robust network design. 
Note that the $\fgc$ problem is the same as the $\onefgc$ problem.
Observe that the problem of finding a minimum-cost $p$-edge connected
spanning subgraph (abbreviated as $\kecss[p]$) corresponds to the
special case of $\pqfgc$ where all edges are safe, and the problem
of finding a minimum-cost $\kecss[(p+q)]$ corresponds to the special
case of $\pqfgc$ where all edges are unsafe.
Informally speaking, the model of $\pqfgc$ interpolates between
$p$-edge connectivity (when all edges are safe) and $(p+q)$-edge
connectivity (when all edges are unsafe).

For each of the problems considered in this paper, any solution
(i.e., output) must be a subgraph of the graph $G$ of the instance;
that is, the (multi) set of edges $F$ of a solution must be a subset
of the (multi) set $E(G)$;
in other words, the number of copies of an edge $e=vw$ in $F$
cannot exceed the number of copies of $e$ in $E(G)$;
see the discussion in \cite[Chapter~3.1]{Schrijver}.

The $\pqfgc$ model is related to the model of Capacitated Network Design.
There are several results pertaining to approximation algorithms for
various problems in Capacitated Network Design; for example, see
Goemans et~al.\ \cite{GoemansGPSTW94} and Chakrabarty
et~al.\ \cite{ChakrabartyCKK15}.
Let $k$ be a positive integer.
The {Capacitated $k$-Connected Subgraph} problem, see
\cite{ChakrabartyCKK15}, is a well studied problem in this area.
We denote this problem by $\capkecss$.  In an instance of
this problem, we have an undirected connected graph $G = (V,E)$,
nonnegative integer edge-capacities $u: E\to\Zp$, nonnegative edge-costs
$c: E\to\Rp$, and a positive integer $k$.  The goal is to find an
edge-set $F \subseteq E$ such that for any nonempty $R\subsetneq{V}$
we have $\sum_{e \in \del_G(R) \cap F} u_e \geq k$, and $c(F)$ is
minimized.
Let $n$ and $m$ denote the number of vertices and edges of $G$, respectively.
For this problem, Goemans et~al.\ \cite{GoemansGPSTW94} give a
$\min(2k,m)$-approximation algorithm,
and Chakrabarty et~al.\ \cite{ChakrabartyCKK15} give a
randomized $O(\log n)$-approximation algorithm.
We mention that for some particular values of $p$ and $q$,
such as $p=1$ (and arbitrary $q$) or $q\leq1$ (and arbitrary $p$),
$\pqfgc$ can be cast as a special case of the $\capkecss$ problem.
The $\fgc$ problem corresponds to the
$\capkecss[2]$ problem where safe edges have capacity~two and
unsafe edges have capacity~one.
Moreover, $\ahmfgc$ corresponds to the $\capkecss[(k+1)]$ problem
where safe edges have capacity~$k+1$ and unsafe edges have capacity~one;
$\ourfgc$ corresponds to the $\capkecss[(k(k+1))]$ problem where
safe edges have capacity~$k+1$ and unsafe edges have capacity~$k$.
We mention that there exist values of $p$ and $q$ (e.g., $p=2, q=2$)
such that the $\pqfgc$ problem differs from the $\capkecss[(p(p+q))]$
problem where safe edges have capacity~$p+q$ and unsafe edges
have capacity~$p$.

\subsection*{Our contributions:} We list our main contributions
and give a brief overview of our results and techniques.

Our first result is based on a simple reduction from $\fgc$ to
the well-known minimum-cost rooted $2$-arborescence problem that
achieves an approximation factor of two for $\fgc$.
This result matches the current best approximation factor known
for the minimum-cost $\twoecss$~problem, and improves on the
$2.527$-approximation algorithm of \cite{AHM21}.
At a high level, our result is based on an extension of the
$2$-approximation algorithm of Khuller and Vishkin \cite{KV94} for
the minimum-cost $\twoecss$~problem.
(In fact, Khuller and Vishkin \cite{KV94} give a reduction
from the minimum-cost $\kecss$~problem to the problem of computing
a minimum-cost rooted $k$-arborescence in a digraph, and
they prove an approximation factor of two for the former problem.)

\begin{theorem} \label{thm:fgc-2apx}
There is a $2$-approximation algorithm for $\fgc$.
\end{theorem}

The following result generalizes Theorem~\ref{thm:fgc-2apx} to the $\ahmfgc$ problem.
Our proof of the generalization of Theorem~\ref{thm:fgc-2apx} is based on a reduction from
$\ahmfgc$ to the minimum-cost rooted $(k+1)$-arborescence problem
(see \cite{Schrijver}, Chapters~52 and~53).  
We lose a factor of $k+1$ in this reduction.

\begin{theorem} \label{thm:kfgc-apx}
There is a $(k+1)$-approximation algorithm for $\ahmfgc$.
\end{theorem}

{
In Section~\ref{sec:capkecss}, we consider the
{Capacitated $k$-Connected Subgraph} problem that we denote by $\capkecss$.
For notational convenience, let $\umax := \max\{u_e : e \in E\}$
denote the maximum capacity of an edge in the given instance of $\capkecss$;
similarly, let ${\umin} := \min\{u_e : e \in E\}$.
Our main result in Section~\ref{sec:capkecss} is
a $\min(k,2\umax)$-approximation algorithm for the $\capkecss$ problem,
stated in Theorem~\ref{thm:capkecss}.
Similar to Theorems~\ref{thm:fgc-2apx}~and~\ref{thm:kfgc-apx}, our proof
of Theorem~\ref{thm:capkecss} is based on a reduction from the
$\capkecss$ problem to the minimum-cost rooted $k$-arborescence problem.
The factor $m$ in the $\min(2k,m)$ approximation factor of Goemans
et~al.\ comes from the fact that a simple greedy strategy yields an
$m$-approximation for the $\capkecss$ problem.
Assuming $\min(k,2\umax) \leq m$, our result is better,
and, in fact, for the standard case of ${\umin}=1,\;\umax=k\ll{m}$,
no previous result achieves an approximation factor of $k$ (to
the best of our knowledge).
Our result above is incomparable to the result in \cite{ChakrabartyCKK15};
our approximation factor is independent of the graph size, whereas
their result is independent of $k$.  The algorithm in
\cite{ChakrabartyCKK15} is probabilistic and its analysis is based
on Chernoff tail bounds.
}

\begin{theorem} \label{thm:capkecss}
There is a $\min(k,2\umax)$-approximation algorithm for the $\capkecss$ problem.
\end{theorem}

{
In Section~\ref{sec:ourfgc}, we present a 4-approximation algorithm
for the $\ourfgc$ problem, see Theorem~\ref{thm:ourfgc}.
Our algorithm in Theorem~\ref{thm:ourfgc} runs in two stages. 
In the first stage we pretend that all edges are safe.
Under this assumption, $\ourfgc$ simplifies to the minimum-cost $\kecss$ problem,
for which several $2$-approximation algorithms are known, see \cite{KV94}, \cite{Jain01}.
We apply one of these algorithms.
Let $H = (V,F)$ be the $\kecss$ found in Stage~1.
In the second stage, our goal is to preserve $k$-edge connectivity
against the failure of any one unsafe edge.
In the graph $H$, consider a cut $\del_H(S)$, $\emptyset\neq{S}\subsetneq{V}$,
that has (exactly) $k$ edges and that contains at least one unsafe edge.
Such a cut, that we call {deficient}, certifies that $F$ is not
feasible for $\ourfgc$, so it needs to be augmented.
The residual problem is that of finding a cheapest augmentation of
all deficient cuts w.r.t.~(with respect to) $F$.
It turns out that this augmentation problem can be formulated as
the minimum-cost $\unfn$-connectivity problem for an uncrossable function $\unfn$
(to be defined in Section~\ref{sec:ourfgc}).
Williamson, Goemans, Mihail and Vazirani \cite{WGMV95} present
a $2$-approximation algorithm for the latter problem.
}

\begin{theorem} \label{thm:ourfgc}
There is a $4$-approximation algorithm for $\ourfgc$.
\end{theorem}

{
In Section~\ref{sec:pqfgc}, we present an $O(q\log{n})$-approximation
algorithm for $\pqfgc$.
As above, our approximation algorithm for $\pqfgc$ runs in two stages. 
In the first stage, we construct an instance of the $\capkecss$
problem (that partially models the given $\pqfgc$~instance), and
then we apply our approximation algorithm for the $\capkecss$ problem
(see Theorem~\ref{thm:capkecss}) to compute a cheap edge-set $F$
that is nearly~feasible for the $\pqfgc$~instance.
We call a cut $\del_G(S)$, $\emptyset\neq S\subsetneq{V}$, {deficient}
(w.r.t.\ $F$) if $|F \cap \del_G(S) \cap \safe| < p$ and $|F \cap
\del_G(S)| < p+q$; thus, a deficient cut is one that certifies the
infeasibility of $F$.
The second stage of our algorithm applies several iterations. In
the first iteration, we find all the deficient cuts of our current
subgraph $H=(V,F)$, and then we apply the greedy algorithm for the
(well-known) hitting-set problem to cover all the deficient cuts.
We repeat such iterations until our current subgraph is a feasible
solution of the $\pqfgc$~instance (i.e., there are no deficient cuts).

\begin{theorem} \label{thm:pqFGC}
There is an $O(q\log{n})$-approximation algorithm for $\pqfgc$.
\end{theorem}
}

{
In Section~\ref{sec:unitfgc}, we consider the unweighted version
of $\fgc$, where each edge has unit cost.  We design an improved
approximation algorithm for this special case.
We give two algorithms for obtaining two candidate solutions to an
instance of unweighted~$\fgc$; the simpler of these algorithms is discussed by
Adjiashvili et~al.\ \cite{AHM20,AHM21}.
Assuming that we have access to an $\alpha$-approximation algorithm
for the minimum-size (i.e., unweighted) $\twoecss$~problem, we argue that the cheaper
of the two candidate solutions is a $\frac{4\alpha}{2\alpha+1}$-approximate
solution to the instance of unweighted~$\fgc$.
We use a result of Seb{\"{o}} \& Vygen \cite{SV14}, and we fix $\alpha=4/3$.

\begin{theorem} \label{thm:unitFGC}
There is a $\frac{16}{11}$-approximation algorithm for unweighted~$\fgc$.
\end{theorem}
}

Section~\ref{sec:ahmfgc} focuses on $\ahmfgc$ and gives our
$(k+1)$-approximation for this problem (Theorem~\ref{thm:kfgc-apx}); the
$2$-approximation for $\fgc$ (Theorem~\ref{thm:fgc-2apx}) follows as a special case.
Section~\ref{sec:capkecss} focuses on the $\capkecss$ problem,
and gives our $\min(k,2\umax)$-approximation
algorithm for this problem (Theorem~\ref{thm:capkecss}).
Section~\ref{sec:ourfgc} focuses on $\ourfgc$, and gives our
$4$-approximation algorithm for this problem (Theorem~\ref{thm:ourfgc}).
Section~\ref{sec:pqfgc} focuses on $\pqfgc$, and gives our
$O(q\log{n})$-approximation algorithm for this problem (Theorem~\ref{thm:pqFGC}).
Section~\ref{sec:unitfgc} focuses on the unweighted version of $\fgc$,
and gives our $\frac{16}{11}$-approximation algorithm for this problem
(Theorem~\ref{thm:unitFGC}).  This section also has an improved
approximation factor for the unweighted version of $\ourfgc$.

}

{ \section{A $(k+1)$-Approximation Algorithm for $\ahmfgc$} \label{sec:ahmfgc}

We give a $(k+1)$-approximation for $\ahmfgc$, where $k$ is a positive integer. 
The $2$-approximation for $\fgc$ (Theorem~\ref{thm:fgc-2apx}) follows as a special case.
Recall that in an instance of $\ahmfgc$ we have an undirected
graph $G = (V,E)$ (with no self loops), a partition of $E$
into safe and unsafe edges, $E=\safe\disjcup\unsafe$, and nonnegative
edge-costs $c: E\to\Rp$.
Our objective is to find a minimum-cost edge-set $F \subseteq E$
such that the subgraph $(V,F)$ remains connected against the failure
of any set of $k$ unsafe edges.

For a subgraph $H$ of $G$ and a vertex-set $S\subseteq{V}$,
we use $\del_{H}(S)$ or $\del_{E(H)}(S)$ to denote
the set of edges in $H$ with exactly one end-vertex in $S$, i.e.,
$\del_{H}(S) := \{ e = uv \in E(H) : |\{u,v\} \cap S| = 1 \}$.  We
drop the subscript when the underlying graph is clear from the
context.

The following characterization of feasible solutions of $\ahmfgc$ is straightforward.

\begin{proposition} \label{pro:cut-prop-ahmfgc}
An edge-set $F \subseteq E$ is feasible for $\ahmfgc$ if and only
if for all nonempty $S \subsetneq V$, the edge-set $F \cap \del(S)$
contains a safe edge or $k+1$ unsafe edges.
{Furthermore, in time polynomial in $n$, we can test if $F$
is feasible for $\ahmfgc$.}
\end{proposition}

We check the feasibility of $F$ for $\ahmfgc$ by creating an auxiliary
capacitated graph that has a capacity of $k+1$ for each safe edge
and a capacity of one for each unsafe edge; then, we test whether
or not the minimum capacity of a cut of the auxiliary graph is at least $k+1$.
{For the rest of this section, we assume that the given
instance of $\ahmfgc$ is feasible.}

As mentioned before, our algorithm for $\ahmfgc$ is based on a
reduction to the minimum-cost rooted $(k+1)$-arborescence problem.
We state a few standard results on arborescences.
Let $D = (W,A)$ be a digraph and let $c': A\to\Rp$ be nonnegative costs on the arcs.
We remark that $D$ may have parallel arcs but it has no self-loops. 
Let $r \in W$ be a designated root vertex. 
For a subgraph $H$ of $D$ and a nonempty vertex-set $S \subsetneq
W$, we use $\din_H(S)$ to denote the set of arcs in $H$ such that
the head of the arc is in $S$ and the tail of the arc is in $W
\setminus S$, i.e.,
$\din_H(S) := \{ a = (u,v) \in A(H) : u \notin S, v \in S\}$.

\begin{definition}[$r$-rooted arborescence] \label{arb-defn}
An $r$-rooted arborescence $(W,T)$ is a subgraph of $D$ satisfying: (i) the undirected version of $T$ is acyclic; and (ii) for every $v \in W \setminus \{r\}$, there is a directed path from $r$ to $v$ in the subgraph $(W,T)$.
\end{definition}

In other words, an $r$-rooted arborescence is a directed spanning tree such that
vertex~$r$ has no incoming arcs and every other vertex has one incoming arc.
An $r$-rooted $k$-arborescence is a union of $k$
arc-disjoint $r$-rooted arborescences.

\begin{definition}[$r$-rooted $k$-arborescence] \label{k-arb-defn}
For a positive integer $k$, a subgraph $(W,T)$ is an $r$-rooted
$k$-arborescence if $T$ can be partitioned into $k$ arc-disjoint
$r$-rooted arborescences.
\end{definition}

The following results on rooted arborescences and the
corresponding optimization problem are useful for us.

\begin{proposition}[\cite{Schrijver}, Chapter~53.8]\label{pro:k-arb-char}
Let $D=(W,A)$ be a digraph, let $r \in W$ be a vertex, and let $k$ be a positive integer. 
Then, $D$ contains an $r$-rooted $k$-arborescence if and only if
$|\din_D(S)|\geq k$ for any nonempty vertex-set $S \subseteq V \setminus \{r\}$.
\end{proposition}

\begin{proposition}[\cite{Schrijver}, Theorem~53.10] \label{pro:min-cost-arb}
In strongly polynomial time, we can obtain an optimal solution to
the minimum-cost $r$-rooted $k$-arborescence problem on $(D,\,c')$,
or conclude that there is no $r$-rooted $k$-arborescence in $D$.
\end{proposition}

\begin{claim} \label{lim-freq}
Let $(W,T)$ be an $r$-rooted $k$-arborescence for an integer $k \geq 1$.
Let $u,v \in W$ be two distinct vertices.
Then, the number of arcs in $T$ that have one end-vertex at $u$ and
the other end-vertex at $v$ (counting multiplicities) is at most $k$.
\end{claim}
\begin{proof}
Since an $r$-rooted $k$-arborescence is a union of $k$ arc-disjoint
$r$-rooted $1$-arborescences, it suffices to prove the result for $k=1$.
The claim holds for $k=1$ because the undirected version of $T$ is acyclic, by definition.
\end{proof}

Informally speaking, our proofs map undirected graphs to their
directed counterparts by bidirecting edges.
We formalize this notion.

\begin{definition}[Bidirected pair] \label{bidirected}
For an undirected edge $e = uv$, we call the arc-set $\{(u,v),(v,u)\}$
a bidirected pair arising from $e$.
\end{definition}

The following lemma shows how a solution $F$ to $\ahmfgc$ can be used
to obtain a rooted $(k+1)$-arborescence (in an appropriate digraph)
of cost at most $(k+1)$ times $c(F)$.

\begin{lemma} \label{fgc-to-arb}
Let $F$ be a $\ahmfgc$ solution.
Consider the digraph $D = (V,A)$ where the arc-set $A$ is defined
as follows: for each unsafe edge $e \in F \cap \unsafe$, we include
a bidirected pair of arcs arising from $e$, and for each safe edge
$e \in F \cap \safe$, we include $k+1$ bidirected pairs arising from $e$.
Consider the natural extension of the cost vector $c$ to $D$ where
the cost of an arc $(u,v) \in A$ is equal to the cost of the edge
in $G$ that gives rise to it.
Then, there is an $r$-rooted $(k+1)$-arborescence in $D$ with cost at most $(k+1)c(F)$.
\end{lemma}
\begin{proof}
Let $(V,T)$ be a minimum-cost $r$-rooted $(k+1)$-arborescence in $D$.
First, we argue that $T$ is well-defined. 
By Proposition~\ref{pro:k-arb-char}, it suffices to show that for any
nonempty $S \subseteq V \setminus \{r\}$, we have $|\din_D(S)| \geq k+1$.
Fix some nonempty $S \subseteq V \setminus \{r\}$.
By feasibility of $F$, $F \cap \del(S)$ contains a safe edge or
$k+1$ unsafe edges (see Proposition~\ref{pro:cut-prop-ahmfgc}).
If $F \cap \del(S)$ contains a safe edge $e = uv$ with $v \in S$,
then by our choice of $A$, $\din_D(S)$ contains $k+1$ $(u,v)$-arcs.
Otherwise, $F \cap \del(S)$ contains $k+1$ unsafe edges, and for
each such unsafe edge $uv$ with $v \in S$, $\din_D(S)$ contains the
arc $(u,v)$.
In both cases we have $|\din_D(S)| \geq k+1$, so $T$ is well-defined. 

We use Claim~\ref{lim-freq} to show that $T$ satisfies the required bound on the cost.
For each unsafe edge $e \in F$, $T$ contains at most $2$ arcs from
the bidirected pair arising from $e$, and for each safe edge $e \in
F$, $T$ contains at most $k+1$ arcs from the (disjoint) union of
$k+1$ bidirected pairs arising from $e$.
Thus, $c(T) \leq 2 \, c(F \cap \unsafe) + (k+1) \, c(F \cap \safe) \leq (k+1) \, c(F)$.
\end{proof}

Lemma~\ref{fgc-to-arb} naturally suggests a reduction from $\ahmfgc$
to the minimum-cost $r$-rooted $(k+1)$-arborescence problem.
We prove the main theorem of this section.

\begin{proofof}{Theorem~\ref{thm:kfgc-apx}}
Fix some vertex $r \in V$ as the root vertex.
Consider the digraph $D = (V,A)$ obtained from $G$ as follows: for
each unsafe edge $e \in \unsafe$, we include a bidirected pair
arising from $e$, and for each safe edge $e \in \safe$, we include
$k+1$ bidirected pairs arising from $e$.
For each edge $e \in E$, let $R(e)$ denote the multi-set of all
arcs in $D$ that arise from $e \in E$.
For any edge $e \in E$ (that could be one of the copies of a multiedge)
and each of the corresponding arcs $\vec{e} \in R(e)$, we define $c_{\vec{e}} := c_e$.
Let $(V,T)$ denote a minimum-cost $r$-rooted $(k+1)$-arborescence in $(D,\,c)$.
By Lemma~\ref{fgc-to-arb}, $c(T) \leq (k+1) c(F^*)$, where $F^*$
denotes an optimal solution to the given instance of $\ahmfgc$.

We finish the proof by arguing that $T$ induces a $\ahmfgc$ solution
$F$ with cost at most $c(T)$.
Let $F := \{ e \in E : R(e) \cap T \neq \emptyset \}$.
By definition of $F$ and our choice of arc-costs in $D$, we have $c(F) \leq c(T)$.
It remains to show that $F$ is feasible for $\ahmfgc$.
Consider a nonempty set $S \subseteq V \setminus \{r\}$.
Since $T$ is an $r$-rooted $(k+1)$-arborescence, by
Proposition~\ref{pro:k-arb-char} we have $|\din_T(S)| \geq k+1$.
If $\din_T(S)$ contains a safe arc (i.e., an arc that arises from
a safe edge), then that safe edge belongs to $F \cap \del(S)$.
Otherwise, $\din_T(S)$ contains some $k+1$ unsafe arcs (that arise from unsafe edges).
Since both orientations of an edge cannot appear in $\din_D(S)$,
we get that $|F \cap \unsafe \cap \del(S)| \geq k+1$.
By Proposition~\ref{pro:cut-prop-ahmfgc}, $F$ is a feasible solution
for the given instance of $\ahmfgc$ with $c(F)\leq(k+1)\cdot\opt$,
where $\opt$ denotes the optimal value of the instance.
\end{proofof}
}

{
\section{The Capacitated $k$-Connected Subgraph Problem} \label{sec:capkecss}

In this section we consider the $\capkecss$ problem. 
We are given a graph $G = (V,E)$,
nonnegative integer edge-capacities $u: E\to\Zp$, nonnegative edge-costs $c: E\to\Rp$,
and a positive integer $k$.
Our goal is to find a spanning subgraph $H = (V,F)$ such that for
all nonempty sets $R \subsetneq V$ we have $u(\del_F(R)) \geq k$,
and the cost $c(F)$ is minimized.

Given an instance of the $\capkecss$ problem, we may assume without
loss of generality that $u_e \in \{1,\dots,k\}$ for all $e \in E$
(we can drop edges with zero capacity and replace edge-capacities $\geq{k+1}$ by $k$).
We also assume that the $\capkecss$~instance is feasible. 
This can be verified in polynomial time by checking if $G$ has any
cut $\del(S)$, $\emptyset\neq{S}\subsetneq{V}$, with capacity $u(\del(S))$ less than $k$.
Let $\umax = \max_{e \in E} u_e$ denote the maximum capacity of an edge in $G$.
Our main result in this section is a $\min(k,2\umax)$-approximation
algorithm for the $\capkecss$ problem (Theorem~\ref{thm:capkecss});
our algorithm is based on a reduction to the minimum-cost rooted $k$-arborescence problem.

\subsection*{Description of our algorithm for the $\capkecss$ problem:}
Let $D=(V,A)$ be the directed graph obtained from $G$ by replacing
every edge $xy \in E$ by $u_{xy}$ pairs of bidirected arcs
$(x,y),(y,x)$, each with the same cost as the edge $xy$; thus, each edge
$e$ in $G$ has $2u_e$ corresponding~arcs in $D$, each of cost $c_e$.
Designate an arbitrary vertex $r \in V$ as the root.
By feasibility of the $\capkecss$~instance, we know that $D$ contains
an $r$-rooted $k$-arborescence (see Proposition~\ref{pro:k-arb-char}).
We use Proposition~\ref{pro:min-cost-arb} on $(D,c)$ to obtain a minimum-cost
$r$-rooted $k$-arborescence $(V,T')$ in polynomial time.
Let $F'$ be the set of all edges $e \in E$ such that at least one of
the $2u_e$ corresponding~arcs in $D$ appears in $T'$.

\begin{lemma} \label{lem:sgc-alg}
The edge-set $F'$ obtained by the above algorithm is feasible for the
given $\capkecss$~instance and it has cost at most $c(T')$.
\end{lemma}
\begin{proof}
Let $R \subsetneq V \setminus \{r\}$ be a
nonempty vertex-set that excludes the root vertex $r$.
Since $(V,T')$ contains $k$ arc-disjoint $r$-rooted arborescences, $|\din_{T'}(R)| \geq k$
(by Proposition~\ref{pro:k-arb-char}).
For each edge $e \in E$,
at most $u_e$ of the corresponding~arcs in $D$ can occur in the arc-set $\din_{T'}(R)$,
by the construction of $D$;
hence, for any edge $e \in \del(R) \cap F'$, $u_e$ is an upper~bound on
the number of corresponding~arcs of $e$ in $\din_{T'}(R)$.
Therefore, $\sum_{e \in \del(R) \cap F'} u_e \geq |\din_{T'}(R)| \geq k$,
and $F'$ is a feasible solution for the $\capkecss$~instance, as required.
For any edge $e \in E$, we only include a single copy of $e$ in $F'$
whenever any of the $2u_e$ corresponding~arcs appear in $T'$, so we
have $c(F') \leq c(T')$.
\end{proof}

We now prove Theorem~\ref{thm:capkecss} by showing that the edge-set $F'$
found by the algorithm has cost $\leq \min(k,2\umax)\cdot\opt$,
where $\opt$ denotes the optimal value of the instance.

\begin{proofof}{Theorem~\ref{thm:capkecss}}
Let $(G(V,E),u,c,k)$ denote a feasible instance of the $\capkecss$ problem.
Let $r \in V$ be the root vertex fixed by the algorithm.
Let $D=(V,A)$ be the digraph and let $(V,T')$ be the $r$-rooted $k$-arborescence
constructed by our algorithm.
Let $F^*$ be an optimal solution to the $\capkecss$~instance, and let $D^*=(V,A^*)$ be
the digraph obtained from $(V,F^*)$ by replacing every edge $xy \in F^*$ by
$u_{xy}$ pairs of bidirected arcs $(x,y),(y,x)$ each with the same cost as edge $xy$.
By feasibility of $F^*$ (for the $\capkecss$~instance),
$D^*$ contains an $r$-rooted $k$-arborescence.
Let $(V,T^*)$ denote an optimal $r$-rooted $k$-arborescence in $D^*$.
Since $D^*$ is a subgraph of $D$ and $(V,T')$ is an optimal $r$-rooted
$k$-arborescence in $D$, we have $c(T') \leq c(T^*)$.
By Lemma~\ref{lem:sgc-alg}, $c(F') \leq c(T')$, so to prove the theorem
it suffices to argue that $c(T^*) \leq \min(k,\,2\umax) c(F^*)$.
To this end, observe that for any edge $e \in F^*$ there are at most
$2u_e$ corresponding~arcs in $A^*$ by construction of $D^*$.
Hence, we have $c(T^*) \leq c(A^*) \leq 2 \umax c(F^*)$.
Next, by definition, $T^*$ can be partitioned into $k$ (arc-disjoint)
$r$-rooted arborescences, each of which can use at most one of the
$2u_e$ corresponding~arcs of an edge $e$ of $G$;
see Claim~\ref{lim-freq}.
It follows that for each edge $e \in F^*$ at most $k$ of the
$2u_e$ corresponding~arcs can appear in $T^*$.
Hence, $c(T^*) \leq k\, c(F^*)$.
This completes the proof.
\end{proofof}
}

{
\section{A $4$-Approximation Algorithm for $\ourfgc$} \label{sec:ourfgc}

Our main result in this section is a $4$-approximation algorithm
for $\ourfgc$ (Theorem~\ref{thm:ourfgc}).
Recall that in an instance of $\ourfgc$, we have a graph $G = (V,E)$,
with a partition of the edge-set into safe and unsafe edges, $E=\safe\cup\unsafe$,
nonnegative edge-costs $c: E\to\Rp$, and a positive integer $k$.
The objective is to find a minimum-cost subgraph that remains
$k$-edge connected against the failure of any one unsafe edge.
We remark that for the $k=1$ case, Theorem~\ref{thm:fgc-2apx} yields a
better approximation factor than Theorem~\ref{thm:ourfgc}.
Let $F^*$ denote an optimal solution to the $\ourfgc$~instance.
The following result pertains to feasible solutions of $\ourfgc$.

\begin{proposition} \label{pro:cut-prop-ourfgc}
An edge-set $F \subseteq E$ is feasible for $\ourfgc$ if and only
if for all nonempty $S \subsetneq V$, the edge-set $F \cap \del(S)$
contains $k$ safe edges or $k+1$ edges.
{Furthermore, in time polynomial in $n$, we can test if $F$
is feasible for $\ourfgc$.}
\end{proposition}
\begin{proof}
The characterization of feasible solutions of $\ourfgc$ follows from the definitions.

We check the feasibility of $F$ for $\ourfgc$ by creating an auxiliary
capacitated graph that has a capacity of $k+1$ for each safe edge
and a capacity of $k$ for each unsafe edge; then, we test whether
or not the minimum capacity of a cut of the auxiliary graph is at least $k(k+1)$.
\end{proof}

{For the rest of this section, we assume that the given
instance of $\ourfgc$ is feasible.}

Proposition~\ref{pro:cut-prop-ourfgc} suggests a two-stage strategy for
finding an approximately optimal solution to $\ourfgc$.
In the first stage, we do not distinguish between safe edges and unsafe edges, and
we compute a cheap $\kecss$  of $G=(V,E)$ that we denote by $H_1 = (V,F_1)$.
Clearly, for every nonempty set $S \subsetneq V$, we have $|\del_{H_1}(S)|\geq{k}$;
if equality holds, then we call $\del_{H_1}(S)$ a \textit{$k$-edge-cut}.
If $F_1$ is feasible for $\ourfgc$, then we are done.
Otherwise, by Proposition~\ref{pro:cut-prop-ourfgc}, the infeasibility
of $F_1$ for $\ourfgc$ is due to $k$-edge-cuts of $H_1$ that contain at
least one unsafe edge. We call such cuts \textit{deficient}.
In the second stage, we address the remaining augmentation problem
for the {deficient} cuts, by casting it as a special case of the
minimum-cost $\unfn$-connectivity problem (defined below).

An instance of the minimum-cost {$\unfn$-connectivity} problem
consists of an undirected graph $G' = (V',E')$, nonnegative edge-costs
$c': E'\to\Rp$, and a requirement function $\unfn : 2^{V'} \to \{0,1\}$
satisfying $\unfn(\emptyset) = \unfn(V) = 0$.
We assume access to $\unfn$ via a value oracle that takes as input
a vertex-set $S \subseteq V$ and outputs $\unfn(S)$.  An edge-set
$F \subseteq E'$ is feasible
if $|F \cap \del_{G'}(S)| \geq \unfn(S)$ for every $S \subseteq V'$.
In other words, $F$ is feasible if and only if for every vertex-set
$S$ with $\unfn(S) = 1$ there is at least one $F$-edge in the cut $\del(S)$.
The objective is to find a feasible $F \subseteq E'$ that minimizes $c'(F)$.
The minimum-cost $\unfn$-connectivity problem can be formulated as an integer program
whose linear relaxation \eqref{eq:primal} is stated below.
For each edge $e \in E'$, the LP (linear program) has a nonnegative variable $x_e$;
informally speaking, $x_e$ quantifies the ``usage'' of the edge $e$ in
a feasible solution to the LP.

\begin{align*} \tag{P} \label{eq:primal}
\quad \min 			& \, \sum_{e \in E'} c'_e x_e 		& &  \\
\text{subject to } 	
	& \, x( \del_{G'}(S) ) \geq \unfn(S) 	& \forall \, S \subseteq V' & & \\
					& \, x_e \geq 0 					& \forall \, e \in E'. & &
\end{align*}

The minimum-cost $\unfn$-connectivity problem has received attention
since it captures many well-known problems in network design.
In particular, it captures the generalized Steiner network
problem. Williamson et~al.\ \cite{WGMV95} gave a primal-dual framework
to obtain approximation algorithms for the minimum-cost $\unfn$-connectivity
problem when $\unfn$ is a  proper function, and more generally,
when $\unfn$ is an uncrossable function (also see the book chapter
by Goemans and Williamson \cite{GW-bookchapter}).

\begin{definition}[Uncrossable function] \label{uncrossable-defn}
A function $\unfn: 2^{V'} \to\{0,1\}$ is called uncrossable
if $\unfn(V') = 0$ and $\unfn$ satisfies the following two conditions:
\begin{enumerate}[(i)]
\item $\unfn$ is symmetric, i.e., $\unfn(S)=\unfn(V' \setminus S)$ for all $S \subseteq V'$;
\item for any two sets $A, B \subseteq V'$ with $\unfn(A) = \unfn(B) = 1$,
either $\unfn(A \cap B) = \unfn(A \cup B) = 1$ or
$\unfn(A \setminus B) = \unfn(B \setminus A) = 1$ holds.
\end{enumerate}
\end{definition}

Under the assumption that \textit{minimal violated sets} can be
computed efficiently throughout, the primal-dual algorithm of
\cite{WGMV95} gives a $2$-approximation for the minimum-cost $\unfn$-connectivity
problem with an uncrossable function $\unfn$.
There is no explicit result in \cite{WGMV95} that can be
quoted verbatim and applied for our purposes, so
we state the most relevant result from \cite{WGMV95}.

\begin{definition}[Minimal violated sets] \label{min-violated-sets}
Let $\unfn: 2^{V'} \to\{0,1\}$ be a requirement function and $F \subseteq E'$ be an edge-set.
A vertex-set $S \subseteq V'$ is said to be violated, w.r.t.\ $\unfn$
and $F$, if $\unfn(S) = 1$ and $F \cap \del_{G'}(S) = \emptyset$.
We say that $S$ is a minimal violated set if
none of the proper subsets of $S$ is violated.
\end{definition}

\begin{proposition}[\cite{WGMV95}, Lemma~2.1] \label{pro:primal-dual}
Let $(G',\, c',\, \unfn)$ be an instance of the minimum-cost
$\unfn$-connectivity problem, where $\unfn : 2^{V'} \to \{0,1\}$
is an uncrossable function that is given via a value oracle.
Suppose that for any $F \subseteq E'$ we can compute all minimal
violated sets (w.r.t.\ $\unfn$ and $F$) in polynomial time. Then,
in polynomial time, we can compute a feasible solution $F\subseteq
E'$ such that $c'(F)\leq{2\,z^*}$, where $z^*$ denotes the optimal
value of the LP~relaxation~\eqref{eq:primal}.
\end{proposition}

We now describe a two-stage algorithm that produces a $4$-approximate
$\ourfgc$ solution in polynomial time, thereby proving
Theorem~\ref{thm:ourfgc}.

\subsection*{Description of our $4$-approximation algorithm for $\ourfgc$:}
Our algorithm runs in two stages.
In the first stage, we construct an instance of the minimum-cost
$\kecss$~problem from the instance of $\ourfgc$, by ignoring the
distinction between the safe edges and the unsafe edges of $G$;
the resulting instance is feasible because $(V,F^*)$ is $k$-edge connected.
Then, we compute a $\kecss$ $H_1 = (V,F_1)$ of $G$ by applying a
2-approximation algorithm to the instance of the minimum-cost
$\kecss$ problem; either the algorithm of Khuller \& Vishkin
\cite{KV94} or the algorithm of Jain \cite{Jain01} could be used.
Clearly, $c(F_1)\leq{2\,c(F^*)}$.
{Next, we compute the collection $\C := \{ S \subseteq V: |\del(S) \cap F_1| = k \}$
of all vertex-sets that correspond to $k$-edge-cuts of $H_1$.}
Consider the requirement function $\unfn : 2^V \to \{0,1\}$ where
\begin{equation}
 	\tag{\ref{sec:ourfgc}:1} \label{eq:def-unfn}
\unfn(S) = 1 \text{ if and only if }
	S \in \C \text{ and } F_1 \cap \del(S) \cap \unsafe \neq \emptyset.
\end{equation}
Consider an instance of the minimum-cost $\unfn$-connectivity problem for the
graph $G' := G - F_1$ with nonnegative edge-costs $c: (E \setminus F_1)\to\Rp$;
note that $F^* \setminus F_1$ is feasible for this instance.
In the second stage, we use Proposition~\ref{pro:primal-dual} to
compute a feasible solution $F_2 \subseteq E \setminus F_1$ for
this instance such that $c(F_2)\leq 2\,c(F^* \setminus F_1)$.
We return the solution $F = F_1 \disjcup F_2$.

We prove Theorem~\ref{thm:ourfgc} via a sequence of lemmas and claims.
Lemma~\ref{lem:uncrossable} below shows that $\unfn$ is uncrossable,
and Claim~\ref{comp-MVS} below shows that we can compute minimal
violated sets (w.r.t.\ $\unfn$ and any $F' \subseteq E \setminus F_1$)
in polynomial time.
These two results together with Proposition~\ref{pro:primal-dual}
imply that our algorithm runs in polynomial time.
Lemma~\ref{ourfgc-feas-cost} shows the correctness of our algorithm,
and proves the approximation factor of four.

\begin{lemma} \label{lem:uncrossable}
$\unfn$ is uncrossable.
\end{lemma}
\begin{proof}
We check that the two properties~(i),~(ii) of an uncrossable function hold for
$\unfn$ (recall Definition~\ref{uncrossable-defn}).
The symmetry of $\unfn$ follows from the symmetry of cuts in undirected graphs.
To check the second property, consider nonempty $A,B \subsetneq V$
satisfying $\unfn(A) = \unfn(B) = 1$.
By definition of $\unfn$, see \eqref{eq:def-unfn}, in the subgraph
$H_1 = (V,F_1)$, both $\del_{F_1}(A)$ and $\del_{F_1}(B)$ are
(minimum) $k$-edge-cuts, and there is at least one unsafe edge in each of these cuts.
Let $e_1$ be an unsafe edge in $\del_{H_1}(A)$ and let $e_2$ be an
unsafe edge in $\del_{H_1}(B)$.
Let $r \in V$ be an arbitrary vertex.
By symmetry of the cut function, we may assume without loss of
generality that $r \notin A \cup B$.
If $A \cap B = \emptyset$, then $\unfn(A \setminus B) = \unfn(B
\setminus A) = 1$, so we are done.
If $A \subseteq B$ or $A \supseteq B$, then $\unfn(A \cap B) =
\unfn(A \cup B) = 1$, so we are done.
Thus, we may assume that $A \cap B, V \setminus (A \cup B), A
\setminus B, B \setminus A$ are all nonempty.
For $S,T\subseteq{V}$, let $E(S,T)$ denote the set of edges of $G$
with exactly one end-vertex in $S$ and exactly one end-vertex in $T$.
By submodularity of the function $d(S) := |\del_{H_1}(S)|$, see \cite{Schrijver}, we have:
\begin{equation}
	\tag{\ref{sec:ourfgc}:2} \label{eq:mincut}
|\del_{H_1}(A \cap B)| = |\del_{H_1}(A \cup B)| = |\del_{H_1}(A \setminus B)| = |\del_{H_1}(B \setminus A)| = k.
\end{equation}
Furthermore, we also have: 
\begin{equation}
	\tag{\ref{sec:ourfgc}:3} \label{eq:crossingedges}
F_1 \cap E(A \setminus B,B \setminus A) = \emptyset \quad \text{ and } \quad F_1 \cap E(A \cap B,V \setminus (A \cup B)) = \emptyset.
\end{equation}
We finish the proof by a case analysis on $e_1$ and $e_2$.
By \eqref{eq:crossingedges}, exactly one of the following cases occurs:
(i)~$e_1 \in E(A \setminus B,V \setminus (A \cup B))$; or
(ii)~$e_1 \in E(A \cap B,B \setminus A)$.
If (i)~occurs, then $\unfn(A \setminus B) = \unfn(A \cup B) = 1$.
Otherwise, (ii)~occurs and $\unfn(A \cap B) = \unfn(B \setminus A) = 1$.
We apply a similar analysis for $e_2$.
Exactly one of the following occurs:
(a)~$e_2 \in E(B \setminus A,V \setminus (A \cup B))$; or
(b)~$e_2 \in E(A \cap B,A \setminus B)$.
If (a)~occurs, then $\unfn(B \setminus A) = \unfn(A \cup B) = 1$.
Otherwise, (b)~occurs and $\unfn(A \cap B) = \unfn(A \setminus B) = 1$.
It is easy to verify that for each of the four combinations, we
either have $\unfn(A \cap B) = \unfn(A \cup B) = 1$ or we have
$\unfn(A \setminus B) = \unfn(B \setminus A) = 1$.
\end{proof}

\begin{claim} \label{comp-MVS}
For any $F' \subseteq E \setminus F_1$, we can compute all minimal
violated sets w.r.t.\ $\unfn$ and $F'$ in polynomial time.
\end{claim}
\begin{proof}
The number of minimum-cuts of a graph on $n$ vertices is (at most)
$O(n^2)$, see \cite{Karger93}, hence, we have $|\C| = O(|V|^2)$.
Using results on network flow algorithms, we can compute $\C$ in
polynomial time, see \cite{Fleischer99}, \cite{NagamochiNI97}.
Since we have explicit access to $\C$, we have a value oracle for $\unfn$.

Let $F' \subseteq E \setminus F_1$ be a given edge-set.
By Definition~\ref{min-violated-sets}, any violated set $S$ w.r.t.\
$F'$ is in $\C$ and has $\unfn(S) = 1$.
We can exhaustively check each of the sets in $\C$ and find each
of the minimal violated sets.
\end{proof}

\begin{lemma} \label{ourfgc-feas-cost}
The edge-set $F=F_1\disjcup{F_2}$ is feasible for $\ourfgc$ and
satisfies $c(F) \leq 4c(F^*)$.
\end{lemma}
\begin{proof}
We first argue that $F$ is feasible.
Since $F_1$ and $F_2$ are edge-disjoint, we have $F \subseteq E$.
We use the characterization of feasible solutions given by Proposition~\ref{pro:cut-prop-ourfgc}.
Consider an arbitrary nonempty vertex-set $S \subsetneq V$.
Since $H_1 = (V,F_1)$ is a $k$-edge connected subgraph of $G$, we
have $|F_1 \cap \del(S)| \geq k$.
 If $|F_1 \cap \del(S)| \geq k+1$, then $|F \cap \del(S)| \geq k+1$. 
 Otherwise, $\del(S)$ is a $k$-edge-cut of $H_1$, i.e., $S \in \C$.
 If $F_1 \cap \del(S)$ contains only safe edges, then $F \cap \del(S)$ contains $k$ safe edges.
Otherwise, by definition of $\unfn$, see \eqref{eq:def-unfn}, $\unfn(S) = 1$.
Next, by feasibility of $F_2$ for the minimum-cost $\unfn$-connectivity problem, we have
$F_2 \cap \del(S) \neq \emptyset$.
Thus, $|F \cap \del(S)| = |F_1 \cap \del(S)| + |F_2 \cap \del(S)| \geq k + 1$.
We show that $c(F) \leq 4\,c(F^*)$ by arguing that each of $c(F_1)$ and
$c(F_2)$ is $\leq 2\,c(F^*)$.
The bound on $c(F_1)$ is immediate from the fact that $F^*$ is
feasible for the instance of the minimum-cost $\kecss$ problem
considered in Stage~1, and by the 2-approximation algorithm used
in Stage~1. Next, by feasibility of $F^* \setminus F_1$ for the
instance of the minimum-cost $\unfn$-connectivity problem,
we have $c(F_2) \leq 2 c(F^* \setminus F_1) \leq 2 c(F^*)$. The lemma follows.
\end{proof}

\bigskip
\noindent
\textbf{Remarks}:
The function $\unfn$ is not a proper function (see \cite{WGMV95}), and
it is not weakly-supermodular (see \cite{Jain01}).
Any proper function $g:2^V\rightarrow\{0,1\}$ must satisfy the maximality property, that is,
$g(A\cup{B})\leq\max\{g(A),g(B)\}$ must hold for any two disjoint sets $A,B\subseteq V$.
Suppose that $k=2$.
Consider the graph $H_1=(V,F_1)$ with $V=\{v_1,v_2,v_3\}$ and with
four unsafe edges, namely, two copies of $v_1v_2$, one copy of $v_1v_3$, and
one copy of $v_2v_3$.
Let $\unfn$ be defined by \eqref{eq:def-unfn}.
Then we have $\unfn(\{v_1\})=0$, $\unfn(\{v_2\})=0$, and $\unfn(\{v_1,v_2\})=1$.
Clearly, maximality is violated by the sets $A=\{v_1\}, B=\{v_2\}$.
Any weakly-supermodular function $g:2^V\rightarrow\Zint$ must satisfy
the property that the inequality
$g(A)+g(B)\leq\max(g(A-B)+g(B-A),\;g(A\cap{B})+g(A\cup{B}))$ holds for
any two sets $A,B\subseteq V$.
Suppose that $k=2$.
Consider the graph $H_1=(V,F_1)$ with $V=\{v_1,v_2,v_3,v_4\}$ and with
one unsafe edge, namely, $v_2v_3$, and five safe edges, namely, two
copies of $v_1v_2$, two copies of $v_3v_4$, and one copy of $v_4v_1$.
Let $\unfn$ be defined by \eqref{eq:def-unfn}.
Let $A=\{v_1,v_2\}$ and let $B=\{v_2,v_3\}$.
Then we have $\unfn(A)=1$, $\unfn(B)=0$,
$\unfn(A-B)=\unfn(\{v_1\})=0$,
$\unfn(B-A)=\unfn(\{v_3\})=0$,
$\unfn(A\cap{B})=\unfn(\{v_2\})=0$,
and
$\unfn(A\cup{B})=\unfn(\{v_1,v_2,v_3\})=0$.
Clearly, the required inequality fails to hold for the sets $A, B$.
}

{
\section{An $O(q\log{n})$-Approximation Algorithm for $\pqfgc$} \label{sec:pqfgc}

In this section, we present an $O(q\log{n})$-approximation algorithm
for $\pqfgc$.
Recall that an instance of $\pqfgc$ consists of an undirected
graph $G = (V,E)$, a partition of $E$ into safe and unsafe
edges, $E = \safe \disjcup \unsafe$,
nonnegative edge-costs $c: E\to\Rp$,
and two integer parameters $p \geq 1$ and $q \geq 0$.
The objective is to find a minimum-cost edge-set $F \subseteq E$
such that the subgraph $(V,F)$ remains $p$-edge~connected against
the failure of any set of at most $q$ unsafe edges, that is,
for any $F'\subseteq\unsafe$ with $|F'|\leq q$,
the subgraph $(V, F\setminus F')$ is $p$-edge~connected.
We assume that $q \geq 2$, since otherwise our results from
Section~\ref{sec:ourfgc} give a $4$-approximation algorithm.
The following result pertains to feasible solutions of $\pqfgc$.

\begin{proposition} \label{pro:cut-prop-pqfgc}
Consider an instance of $\pqfgc$.
An edge-set $F \subseteq E$ is feasible if and only if for all
nonempty $S \subsetneq V$, the edge-set $F \cap \del(S)$ contains
$p$ safe edges or $p+q$ edges.
{Furthermore, in time polynomial in $n$, we can test if $F$
is feasible for $\pqfgc$.}
\end{proposition}
\begin{proof}
The characterization of feasible solutions of $\pqfgc$ follows from the definitions.

We check the feasibility of $F$ for $\pqfgc$ by creating an auxiliary
capacitated graph that has $G = (V,F)$ as the underlying graph, and
that has a capacity of $p+q$ for each safe edge and a capacity of
$p$ for each unsafe edge.
Then, we compute a minimum-capacity cut of the auxiliary graph,
call it $\del_F(S^*)$; note that $S^*$ is nonempty and $S^*\subsetneq{V}$.
Let $\mu$ denote the capacity of $\del_F(S^*)$;
thus, $\mu = (p+q)\,|\del_F(S^*)\cap\safe| + p\,|\del_F(S^*)\cap\unsafe|$.
If $\mu < p(p+q)$, then $F$ is infeasible.
Otherwise, we compute the set $\widehat{\C}$ of all cuts of the auxiliary graph
that have capacity between $\mu$ and $2\mu$, by applying
the polynomial-time algorithm of Nagamochi, Nishimura and Ibaraki
\cite{NagamochiNI97} that enumerates over all $2$-approximate
minimum-cuts of a capacitated graph.
We exhaustively check whether or not each of the cuts in $\widehat{\C}$
has either $(p+q)$ edges or has $p$ safe edges.
Clearly, $F$ is infeasible if $\widehat{\C}$ contains a cut that violates this condition.
Otherwise, $F$ is feasible because any cut $\del_F(S)$, $\emptyset\neq{S}\subsetneq{V}$, 
of the auxiliary graph that is not
in $\widehat{\C}$ has capacity $\geq2\,\mu\ge2\,p(p+q)$, and so either
$(p+q)\,|\del_F(S)\cap\safe| \geq p(p+q)$,
that is, $\del_F(S)$ has $\geq{p}$ safe edges, or
$p\,|\del_F(S)\cap\unsafe| \geq p(p+q)$,
that is, $\del_F(S)$ has $\geq{p+q}$ (unsafe) edges.
\end{proof}

For the rest of this section, we assume that the given instance of $\pqfgc$ is feasible.
Let $F^*$ denote an optimal solution for the $\pqfgc$~instance,
and let $\opt = c(F^*)$ denote the optimal value.

Given an edge-set $F$ (i.e., a candidate solution), we call a cut
$\del(S)$, $\emptyset\neq S\subsetneq{V}$, \textit{deficient}
if $|F \cap \del(S) \cap \safe| < p$ and $|F \cap \del(S)| < p+q$;
thus, a deficient cut is one that certifies the infeasibility of $F$.

First, we give an overview of our $O(q \log n)$-approximation
algorithm for $\pqfgc$.
Our algorithm runs in two stages. 
In the first stage, we construct an instance of the $\capkecss$
problem (that partially models the given $\pqfgc$~instance), and
then we apply our approximation algorithm for the $\capkecss$ problem
(see Theorem~\ref{thm:capkecss}) to compute an edge-set $F$ of cost $O(q \cdot \opt)$
that is ``nearly~feasible'' for the $\pqfgc$~instance.
In more detail, 
the set of cuts that are deficient w.r.t.\ $F$ has size polynomial in $n$, and
the set is computable in time polynomial in $n$.
The second stage of our algorithm applies several iterations.
In each iteration $\ell=1,2,\dots$, we find all the deficient cuts of our current
subgraph $H=(V,F)$ and then we apply the greedy algorithm for the
(well-known) hitting-set problem to
find an edge-set $F_{\ell}$ that covers all the deficient cuts
(i.e., each of the deficient cuts contains at least one edge of $F_{\ell}$).
Then, we update $F$ to $F\disjcup F_{\ell}$, i.e., we augment $F$ by the edge-set
computed by the greedy algorithm, and then
we re-compute the set of deficient cuts w.r.t.\ the updated subgraph $H=(V,F)$.
We stop iterating when there are no deficient cuts w.r.t.\ $H=(V,F)$;
thus, at the termination of the second stage, $F$ is feasible for the $\pqfgc$~instance.
The greedy algorithm (for the hitting-set~instances that arise) achieves an
approximation factor of $O(\log n)$.

{
We discuss the hitting-set problem and state the approximation factor
of the greedy algorithm for this problem.

\begin{definition}[Hitting-Set Problem] \label{defn:hittingset}
Given a ground~set $\E=\{e_1,\dots,e_{\hat{n}}\}$ of elements, a
collection $\Rc = \{R_1,\dots,R_{\hat{m}}\}$ of subsets of $\E$, and
nonnegative costs $c: \E\to\Rp$ on the elements of $\E$, find a
minimum-cost subset $F \subseteq \E$ such that for every $i\in\{1,\dots,|\Rc|\}$,
we have $F \cap R_i \neq \emptyset$.
\end{definition}

It is well-known that there is an $O(\log{|\Rc|})$-approximation algorithm
for the hitting-set problem based on the greedy strategy, see \cite{Vazirani-book}.

\begin{proposition}[\cite{Vazirani-book}] \label{pro:hittingsetapx}
There is an $O(\log{|\Rc|})$-approximation algorithm for the hitting-set
problem that runs in time polynomial in $|\E|$ and $|\Rc|$.
\end{proposition}
}

\subsection*{Constructing an instance of $\capkecss$:}
Given an instance of $\pqfgc$, we construct an instance of
the $\capkecss$ problem on the underlying graph $G = (V,E)$ with
edge costs $c: E\to\Rp$.

We consider two cases, one for $p>q$, and the other for $p\le{q}$, and
we use different values of the edge capacities for the two cases.
In the first case, we use unit edge capacities, and in the second case,
we use the edge capacities given in the proof of Proposition~\ref{pro:cut-prop-pqfgc}.
Let us explain our choice of edge capacities.
Recall that our plan is to apply the
$\min(k,2\umax)$-approximation algorithm for $\capkecss$
(Theorem~\ref{thm:capkecss}) to compute an edge-set $F$ of cost
$O(q\cdot\opt)$ that is ``nearly~feasible'' for the $\pqfgc$~instance.
In the first case ($p>q$, unit edge capacities),
the edge-set $F$ (computed via Theorem~\ref{thm:capkecss})
has cost $\leq 2\cdot\opt=O(q\cdot\opt)$.
In the second case ($p\le{q}$, edge capacities of $p$ or $p+q$),
the edge-set $F$ (computed via Theorem~\ref{thm:capkecss})
has cost $\leq 2(p+q)\cdot\opt=O(q\cdot\opt)$.

\bigskip

\begin{description}
\item[Case~1 $p>q$]
$k:=p$, and all edges have unit capacity i.e., $u_e := 1$ for all $e\in{E}$.
\bigskip

\item[Case~2 $p\le{q}$]
$k:=p(p+q)$, each safe edge has capacity $(p+q)$, and
                 each unsafe edge has capacity $p$, that is,
$u_e = p+q$ if $e\in\safe$, and
$u_e = p  $ if $e\in\unsafe$.
\end{description}
\bigskip

Let $\umax = \max_{e \in E} u_e$ be the maximum edge capacity.

First, we argue that the instance of $\capkecss$ is feasible.
Recall that $F^*$ denotes an optimal solution for the given (feasible) $\pqfgc$~instance.
By Proposition~\ref{pro:cut-prop-pqfgc}, for any nonempty set $S
\subsetneq V$, we either have $|\del(S) \cap F^* \cap \safe| \geq p$ or
$|\del(S) \cap F^*| \geq p+q$.
The feasibility of the $\capkecss$~instance follows from showing
that for all nonempty sets $S\subsetneq{V}$, we have $u(\del(S)) \geq k$.
To this end, fix such an $S$ and recall the choice of edge capacities
that depends on the relative values of $p$ and $q$.
If $p > q$, then 
\begin{equation}
	\tag{\ref{sec:pqfgc}:1} \label{eq:cutcap1}
u(\del(S)) = |\del(S)| \geq |\del(S) \cap F^*| \geq p = k.
\end{equation}
On the other hand, if $p \leq q$, then we have:
\begin{equation}
	\tag{\ref{sec:pqfgc}:2} \label{eq:cutcap2}
u(\del(S)) \geq u(\del(S) \cap F^*) \geq
\max((p+q) \cdot |\del(S) \cap F^* \cap \safe|,\: p \cdot |\del(S) \cap F^*|) \geq
p(p+q) = k.
\end{equation}

Let $F\subseteq{E}$ be a feasible solution to the $\capkecss$~instance;
thus, every nonempty $S\subsetneq V$ has $u(\del(S) \cap F)\geq k$.
Let 
\[
\C := \{ S \subsetneq V : S \neq \emptyset, u(\del(S) \cap F) \leq 2k, |\del(S) \cap F| < p+q, |\del(S) \cap F \cap \safe| < p \}.
\]
Informally speaking, in the capacitated graph $H = (V,\, F,\, u)$ that
corresponds to the edge-set $F$,
$\C$ is the collection of all vertex-sets that correspond to
2-approximate minimum-cuts that violate the feasibility requirement
stated in Proposition~\ref{pro:cut-prop-pqfgc}.

\subsection*{Description of our two-stage algorithm for $\pqfgc$:}
In the first stage of the algorithm, we use Theorem~\ref{thm:capkecss}
to obtain a feasible solution $F$ for the above $\capkecss$~instance
such that the cost $c(F)$ is $\leq \min(k,2\umax)\cdot\opt$.
The second stage of the algorithm consists of several iterations
that augment edges to $F$ until all the deficient cuts w.r.t.\ $F$
have been fixed.
The $\ell^\mathrm{th}$ iteration (for some $\ell=1,2,\dots$)
consists of solving an instance of the hitting-set problem: we want
to hit all sets in the collection $\{ \del(S) \cap (E \setminus F)~:~S\in\C\}$
by using edge-elements $e \in E \setminus F$, where
the cost of $e$ is $c_e$ (in the hitting-set instance).
Let $F'_\ell$ denote a hitting-set computed by the greedy algorithm for
the above hitting-set~instance.
We update $F$ to $F \disjcup F'_\ell$, and recompute $\C$ using the new $F$.
As long as $\C$ is not empty, we repeat the above iteration.
When $\C$ becomes empty, we return the current $F$ as a feasible
solution of the given $\pqfgc$~instance.
Assuming that the number of iterations in the second stage is $O(q)$,
the cost of $F$ is $O(q\cdot \opt + O(q\log{n})\cdot \opt) = O(q\log{n})\cdot\opt$.

Observe that each of the hitting-set~instances constructed by the
algorithm is feasible, because for any deficient cut $\del(S)$
w.r.t.\ the current solution $F$, we have
  $(F^* \setminus F) \cap \del(S) \neq \emptyset$, that is,
$F^* \setminus F$ is a feasible hitting-set.

The next result shows that the algorithm finds a feasible solution.

\begin{lemma} \label{lem:pqfgc-correct}
The edge-set $F$ returned by the algorithm is feasible
for the given $\pqfgc$~instance.
\end{lemma}
\begin{proof}
By the design of the second stage of the algorithm, the solution
$F$ is repeatedly augmented as long as $\C$ is nonempty, so it
suffices to argue that every deficient cut (w.r.t.\ the current $F$)
belongs to $\C$.
Let $\del(S)$, $\emptyset\neq{S}\subsetneq{V}$,  be an arbitrary deficient cut w.r.t.\ $F$.
By definition, $|\del(S) \cap F \cap \safe| < p$ and $|\del(S) \cap F| < p+q$.
To show that $S$ belongs to $\C$, we need to show that $u(\del(S) \cap F) \leq 2k$ holds. 
If $p > q$, then 
\begin{equation}
	\tag{\ref{sec:pqfgc}:3} \label{eq:defcutcap1}
u(\del(S) \cap F) = |\del(S) \cap F| < p+q < 2p = 2k.
\end{equation}
On the other hand, if $p \leq q$, then 
\begin{equation}
	\tag{\ref{sec:pqfgc}:4} \label{eq:defcutcap2}
u(\del(S) \cap F) = p \cdot |\del(S) \cap F| + q \cdot |\del(S) \cap F \cap \safe| <
	p(p+q) + pq < 2p(p+q) = 2k.
\end{equation}
Thus, in either case, deficient cuts are 2-approximate minimum-cuts
in the capacitated graph $H=(V,F,u)$, and they belong to $\C$.
\end{proof}

\medskip

We complete the proof of Theorem~\ref{thm:pqFGC} by arguing that
the above algorithm finds a feasible solution of the $\pqfgc$~instance
of cost $O(q\log{n})\cdot \opt$ in polynomial time.

{
\begin{lemma} \label{lem:pqfgc-apx}
The above two-stage algorithm runs in time polynomial in $n$ and
returns a feasible solution $F\subseteq{E}$ for $\pqfgc$
such that $c(F)\leq O(q\log n)\cdot \opt$.
\end{lemma}
\begin{proof}
We argue that the output of the algorithm has cost at most $O(q \log n) \cdot \opt$.
It is easy to see that the cost of the first-stage solution is $O(q)
\cdot \opt$ because of the approximation factor from
Theorem~\ref{thm:capkecss}; as mentioned before, $F^*$ is feasible
for the $\capkecss$~instance, and $\min(k,\umax) \leq 2q$, 
for all relevant values of $p$ and $q$.

We bound the cost incurred in the second stage by arguing that:
(i)~each iteration leads to an additional cost of at most
$O(\log{n}) \cdot \opt$; and (ii)~there are at most $q$ iterations.

In the capacitated graph $H(V,F,u)$, all deficient cuts are
2-approximate minimum-cuts, i.e., the capacity of each deficient
cut is at most two times the capacity of a minimum-cut.
Karger's bound \cite{Karger93} on the number of $2$-approximate
minimum-cuts implies that $|\C| = O(n^4)$.
By using the algorithm of Nagamochi et~al.~\cite{NagamochiNI97}
we can explicitly compute the collection $\C$ in (deterministic) polynomial time.
Since $F^* \setminus F$ is a feasible solution to each of the hitting-set~instances
constructed in the second stage,
Proposition~\ref{pro:hittingsetapx} implies that
the cost of the augmenting edge-set found in each iteration is
$O(\log |\C|) \cdot c(F^* \setminus F) = O(\log n) \cdot \opt$,
thereby showing~(i).

Next, we bound the number of iterations via a case analysis. 
First, suppose that $p > q$. Then each iteration increases the
capacity of a deficient cut by at least one. By \eqref{eq:cutcap1},
every deficient cut has capacity at least $p$ at the end of the
first stage, and due to \eqref{eq:defcutcap1}, a cut is no longer
deficient once its capacity is at least $p+q$.
Next, suppose that $p\leq q$.
We use a similar argument for this case.
Now, each iteration increases the capacity of a deficient cut
by at least $p$. By \eqref{eq:cutcap2}, every deficient cut has
capacity at least $p(p+q)$ at the end of the first stage, and due to
\eqref{eq:defcutcap2}, a cut is longer deficient once its capacity
is at least $p(p+q) + pq$.
Overall, we have $c(F) \leq (2q + q \cdot O(\log n)) \cdot \opt =
O(q \log n) \cdot \opt$, as desired.

Lastly, we argue that the entire algorithm runs in polynomial time. 
The first stage runs in (deterministic) polynomial time,
by Theorem~\ref{thm:capkecss}.
The second stage also runs in (deterministic) polynomial time because
the number of iterations is at most $q$ ($\leq n$), and in each
iteration we solve a polynomial-sized hitting-set~instance.
This completes the proof of the lemma.
\end{proof}
}
}

{
\section{Unweighted problems: $\onefgc$ and $\ourfgc$} \label{sec:unitfgc}

Consider the unweighted version of $\fgc$ where each edge has unit cost, i.e.,
$c_e = 1$ for all $e \in E$. 
We present a $\frac{16}{11}$-approximation algorithm (see Theorem~\ref{thm:unitFGC}).
To the best of our knowledge, this is the first result that 
improves on the approximation factor of $\frac32$ for unweighted~$\fgc$.
In fact, we give two algorithms for obtaining two candidate solutions to an
instance of unweighted~$\fgc$; the simpler of these algorithms is discussed by
Adjiashvili et~al.\ \cite{AHM20,AHM21}.
Assuming that we have access to an $\alpha$-approximation algorithm
for the minimum-size (i.e., unweighted) $\twoecss$~problem, we argue that the cheaper
of the two candidate solutions is a $\frac{4\alpha}{2\alpha+1}$-approximate
solution to the instance of unweighted~$\fgc$.
Adjiashvili et~al.\ \cite{AHM21} gave an
$\bigl(\frac{\alpha}{2} + 1\bigr)$-approximation algorithm for
unweighted~$\fgc$, assuming the existence of an $\alpha$-approximation
algorithm for the minimum-size (i.e., unweighted) $\twoecss$~problem;
this implies a $\frac53$-approximation algorithm for unweighted~$\fgc$
by using a result of Seb{\"{o}} and Vygen \cite{SV14}.
The algorithm in \cite{AHM21} starts with a maximal forest of safe edges in the graph.
At the end of this section, we give an example showing that
the (asymptotic) approximation factor achievable by such an algorithm is at least $\frac{3}{2}$. 
Our main result in this section is the following.

\begin{theorem} \label{thm:FGCto2ECSS}
Suppose that there is an $\alpha$-approximation algorithm for the
minimum-size (i.e., unweighted) $\twoecss$~problem.
Then, there is a $\frac{4\alpha}{2\alpha+1}$-approximation algorithm for unweighted~$\fgc$.
\end{theorem}

Theorem~\ref{thm:unitFGC} follows from the above theorem by using
the $\frac43$-approximation algorithm of Seb{\"{o}} and Vygen
\cite{SV14} for the minimum-size $\twoecss$~problem.
Before delving into the proof of Theorem~\ref{thm:FGCto2ECSS}, we
introduce some basic results on $\tjn$-joins, which will be useful
in our algorithm and its analysis.
Let $G' = (V',E')$ be an undirected graph with no self-loops
and let $c': E'\to\Rp$ be nonnegative costs on the edges.

\begin{definition}[$\tjn$-join] \label{join-defn}
Let $\tjn \subseteq V'$ be a subset of vertices with $|\tjn|$ even.
A subset $J \subseteq E'$ of edges is called a $\tjn$-join if $\tjn$ is equal to the set of vertices of odd degree in the subgraph $(V',J)$.
\end{definition}

The following classical result on finding a minimum-cost $\tjn$-join is due to Edmonds.

\begin{proposition}[\cite{Schrijver}, Theorem~29.1] \label{pro:min-cost-join}
Given $(G',\,c')$, we can either obtain a minimum-cost $\tjn$-join, or
conclude that there is no $\tjn$-join, in strongly polynomial time. 
\end{proposition}

The $\tjn$-join polytope is the convex hull of the incidence vectors of $\tjn$-joins.
Edmonds \& Johnson showed that
the dominant of the $\tjn$-join polytope has a simple linear description.

\begin{proposition}[\cite{Schrijver}, Corollary~29.2b] \label{pro:join-dominant}
The dominant of the $\tjn$-join polytope is given by $\{ x \in \Rp^{E'} : x(\del_{G'}(S)) \geq 1 \, \forall \, S \subsetneq V' \text{ s.t. } |S \cap \tjn| \text{ odd} \}$.
\end{proposition}

Consider an instance of unweighted~$\fgc$ consisting of a graph
$G = (V,E)$ with a specified partition of $E$ into a set of safe
edges and a set of unsafe edges, $E=\safe\disjcup\unsafe$.
We will assume that $G$ is connected and has no unsafe bridges,
since otherwise the instance is infeasible.
Let $F^*$ denote an optimal solution.

\subsection*{Join-based algorithm for unweighted $\fgc$:}
Let $T$ be a spanning tree in $G$ that maximizes the number of safe edges.
Clearly, for each safe edge $e=uv$ in $E(G)-T$, the (unique)
${u,v}$-path in $T$ consists of safe edges; hence, the graph obtained
from $G$ by contracting all the safe edges of $T$ has no safe edges.
If $|T \cap \safe| = |V|-1$, then $T$ is an optimal $\fgc$ solution
for the given instance, and we are done.
Otherwise, let $T' := T \cap \unsafe$ be the (nonempty) set of unsafe edges in $T$.
Let $G' = (V',E')$ denote the graph obtained from $G$ by contracting all the (safe) edges
in $T \setminus T'$.
(We remove all self-loops from $G'$, but retain
parallel edges that arise due to edge contractions.)
Note that all edges in $E'$ are unsafe (by the discussion above),
and $T'$ is a spanning tree of $G'$.
Let $W'$ denote the (nonempty) set of odd degree vertices in the subgraph $(V',T')$.
Using Proposition~\ref{pro:min-cost-join}, in polynomial time, we compute
a minimum-cardinality $W'$-join in $G'$, and denote it by $J' \subseteq E'$.
By our choice, the subgraph $(V',T' \disjcup J')$ is connected and
Eulerian, so it is $2$-edge connected in $G'$.
Consider the multiset $F_1 = T \disjcup J'$ consisting of edges in
$E$; if an edge $e$ appears in both $T'$ and $J'$, then we include
two copies of $e$ in $F_1$.

\noindent If $F_1$ contains at most one copy of each edge in $E$, then $F_1$ is $\fgc$-feasible.
Otherwise, we modify $F_1$ to get rid of all duplicates without increasing $|F_1|$.
Consider an unsafe edge $e \in E'$ that appears twice in $F_1$, i.e., $e$ belongs to both $T'$ and $J'$.
We remove a copy of $e$ from $F_1$.
If this does not violate $\fgc$-feasibility, then we take no further action.
Otherwise, the second copy of $e$ in $F_1$ is an unsafe bridge in
$(V,F_1)$ that induces a cut $\del(S)$,
$\emptyset\neq{S}\subsetneq{V}$, in $G$.  By our assumption that $G$
has no unsafe bridges, there is another edge $e' \in E$ that is in
$\del(S)$ but not in $F_1$.
We include $e'$ in $F_1$.
This finishes the description of our first algorithm.

\smallskip
At the end of the de-duplication step, $F_1$ is $\fgc$-feasible and it contains at most one copy of any edge $e \in E$. 
It is also clear that $|F_1| \leq |T|+|J'|$.
The following claim gives a bound on the quality of our first solution.

\begin{claim} \label{maxsafe}
We have $|J'| \leq \frac12 |F^* \cap \unsafe|$. Hence, $|F_1| \leq |F^* \cap \safe| + \frac32 |F^* \cap \unsafe|$.
\end{claim}
\begin{proof}
We prove the claim by constructing a fractional $W'$-join of small size.
Recall that we chose $T$ such that $|T \cap \safe|$ is maximum,
and we obtained $G'$ by contracting connected components in $(V,T \setminus T')$.
$G'$ consists of only unsafe edges, and moreover, $G'$ is $2$-edge connected 
because $G$ has no unsafe bridges (by our assumption).
Let $B := F^* \cap E'$ denote the set of unsafe edges in the optimal
solution $F^*$ that also belong to $G'$.
Consider the vector $z := \frac12 \chi^B$ where $\chi^B \in [0,1]^{E'}$
is the incidence vector of $B$ in $G'$.
Let $\del(S')$, $\emptyset\neq{S'}\subsetneq{V(G')}$, be an arbitrary
cut in $G'$ and let $\del(S)$ be the unique cut in $G$ that gives rise
to $\del(S')$ when we contract (safe) edges in $T \setminus T'$.
Since $F^*$ is $\fgc$-feasible and there are no safe edges in
$\del_G(S)$, we must have $|B \cap \del_{G'}(S')| \geq 2$.
Consequently, $z(\del_{G'}(S')) = \frac12 |B \cap \del_{G'}(S')| \geq 1$. 
By Proposition~\ref{pro:join-dominant}, $z$ lies in the dominant of the
$W'$-join polytope, i.e., $z$ dominates a fractional $W'$-join.
Since $J'$ is a min-cardinality $W'$-join,
$|J'| \leq \mathbf{1}^T z \leq \frac12 |F^* \cap \unsafe|$.
We bound the size of $F_1$ by using the trivial bound $|T| \leq |F^*|$: 
\[
|F_1| \leq |F^*| + |J'| \leq |F^* \cap \safe| + \frac32 |F^* \cap \unsafe|. \qedhere
\]
\end{proof}

Our second algorithm uses the $\alpha$-approximation for the
minimum-size $\twoecss$~problem as a subroutine.
Informally speaking, the solution returned by this algorithm has
the property that its size complements that of $F_1$.

\subsection*{$\twoecss$-based algorithm for unweighted $\fgc$:}
Consider the graph $G''$ obtained from $G$ by duplicating every safe edge in $E$.
Similarly, let $F''$ be the multiedge-set obtained from $F^*$ by
duplicating every safe edge in $F^*$.
Clearly, $(V,F'')$ is a $2$-edge connected subgraph of $G''$
consisting of $2|F^* \cap \safe| + |F^* \cap \unsafe|$ edges.
Let $F_2$ be the output of running the $\alpha$-approximation
algorithm for the minimum-size $\twoecss$~problem on $G''$.
Since $F_2$ is $2$-edge connected and only safe edges can appear more than once in $F_2$
(because $G''$ only has duplicates of safe edges),
we can drop the extra copy of all safe edges while maintaining $\fgc$-feasibility in $G$.
This finishes the description of our second algorithm.

\smallskip

The following claim is immediate.

\begin{claim} \label{doublesafe}
We have $|F_2| \leq 2 \alpha |F^* \cap \safe| + \alpha |F^* \cap \unsafe|$.
\end{claim}

We end this section with the proof of our main result on unweighted~$\fgc$.

\begin{proofof}{Theorem~\ref{thm:FGCto2ECSS}}
Given an instance of unweighted~$\fgc$, we compute two candidate solutions $F_1$ and $F_2$
as given by the two algorithms described above.
The solution $F_1$ can be computed using algorithms for the
$\mst$~problem and the minimum-weight $W'$-join problem,
followed by basic graph operations.
The solution $F_2$ can be computed using the given $\alpha$-approximation
algorithm for the minimum-size $\twoecss$~problem.
We show that the smaller of $F_1$ and $F_2$ is a
$\frac{4\alpha}{2\alpha+1}$-approximate solution
for the instance of unweighted~$\fgc$.
By Claims~\ref{maxsafe}~and~\ref{doublesafe}: 
\[
\min(|F_1|,\, |F_2|) \leq \frac{2\alpha}{2\alpha+1} |F_1| + \frac{1}{2\alpha+1} |F_2| = \frac{4\alpha}{2\alpha+1} |F^*| \qedhere
\]
\end{proofof}

As mentioned earlier, we have an example (see
Figure~\ref{fig:3/2tight-instance} below) such that any algorithm for
unweighted~$\fgc$ that starts with a maximal forest on safe edges
achieves an (asymptotic) approximation factor of $\frac{3}{2}$ or more.

{
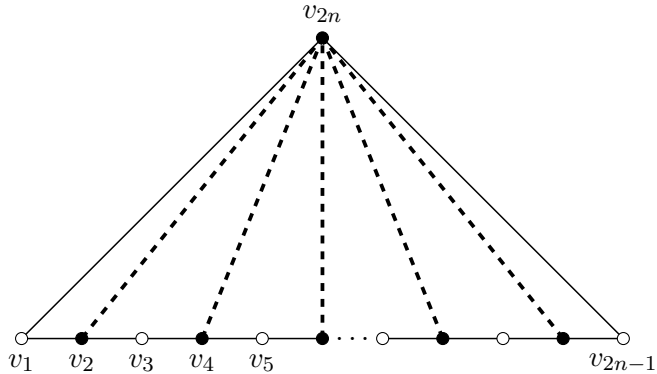
\begin{figure}[htbp]
\centering    
\begin{tikzpicture}[scale=0.8]
	\draw [-] [black, line width=0.2mm] plot coordinates {(6,0) (10,0)};
	\draw [-] [black, line width=0.2mm] plot coordinates {(0,0) (5,0)};
	\draw [-] [black, line width=0.2mm] plot coordinates {(0,0) (5,5)};
	\draw [-] [black, line width=0.2mm] plot coordinates {(10,0) (5,5)};

	\draw [dashed] [black, line width=0.5mm] plot coordinates {(1,0) (5,5)};
	\draw [dashed] [black, line width=0.5mm] plot coordinates {(3,0) (5,5)};
	\draw [dashed] [black, line width=0.5mm] plot coordinates {(5,0) (5,5)};
	\draw [dashed] [black, line width=0.5mm] plot coordinates {(7,0) (5,5)};
	\draw [dashed] [black, line width=0.5mm] plot coordinates {(9,0) (5,5)};
	\node[text = black,fill=white] (label) at (5.5,0) {$\ldots$};	

	\node[text = black,fill=white] (label) at (0,-0.4) {$v_1$};	
	\node[text = black,fill=white] (label) at (1,-0.4) {$v_2$};	
	\node[text = black,fill=white] (label) at (2,-0.4) {$v_3$};	
	\node[text = black,fill=white] (label) at (3,-0.4) {$v_4$};
	\node[text = black,fill=white] (label) at (4,-0.4) {$v_5$};	
	\node[text = black,fill=white] (label) at (10,-0.4) {$v_{2n-1}$};	
	\node[text = black,fill=white] (label) at (5,5.4) {$v_{2n}$};	

	\draw[fill=white] (0,0) ellipse (0.1cm and 0.1cm);
	\draw[fill=black] (1,0) ellipse (0.1cm and 0.1cm);
	\draw[fill=white] (2,0) ellipse (0.1cm and 0.1cm);
	\draw[fill=black] (3,0) ellipse (0.1cm and 0.1cm);
	\draw[fill=white] (4,0) ellipse (0.1cm and 0.1cm);
	\draw[fill=black] (5,0) ellipse (0.1cm and 0.1cm);
	\draw[fill=white] (6,0) ellipse (0.1cm and 0.1cm);
	\draw[fill=black] (7,0) ellipse (0.1cm and 0.1cm);
	\draw[fill=white] (8,0) ellipse (0.1cm and 0.1cm);
	\draw[fill=black] (9,0) ellipse (0.1cm and 0.1cm);
	\draw[fill=white] (10,0) ellipse (0.1cm and 0.1cm);
	\draw[fill=black] (5,5) ellipse (0.1cm and 0.1cm);
\end{tikzpicture}

    \caption{In this instance we have a graph on $2n$ vertices. The set
    of unsafe edges, shown using solid lines, forms a Hamiltonian cycle.
    For each $i=1,\dots,n-1$, there is a safe edge, shown using a thick
    dashed line, between $v_{2i}$ and $v_{2n}$. The solution consisting
    of all unsafe edges is feasible, and any feasible solution must
    contain all unsafe edges, so the value of an optimal integral solution is $2n$.
    Any feasible solution
    that contains a maximal forest on safe edges has size at least $3n-1$.}
    \label{fig:3/2tight-instance}
\end{figure}
}

\pagebreak

\begin{minipage}{\textwidth}
{
\subsection*{Improved approximation factor for unweighted $\ourfgc$:}
Finally, we focus on unweighted~$\ourfgc$ where each edge has unit cost.
We can improve on the approximation factor of four of Theorem~\ref{thm:ourfgc},
by using the same method (see Section~\ref{sec:ourfgc}),
except that in the first stage
we apply the best approximation algorithm known for
the minimum-size (unweighted) $\kecss$ problem.
Let $\alpha_k$ denote the best approximation factor known for the latter problem.
Note that $\alpha_2=4/3$ (see Seb{\"{o}} \& Vygen \cite{SV14}),
$\alpha_3=1.5$ (see Gabow \cite{Gabow04}), and, in general,
$\alpha_k < 1 + \frac{1.91}{k}$ (see Gabow \& Gallagher \cite{GabowGallagher12}).

\bigskip

\begin{proposition}
There is a $(2+\alpha_k)$-approximation algorithm for unweighted~$\ourfgc$.
Thus, the approximation factor is
$\frac{10}{3}$ for $k=2$,
$\frac{7}{2}$ for $k=3$, and
it is less than $(3 + \frac{1.91}{k})$ for all integers $k\ge4$.
\end{proposition}
}
\end{minipage}
}
\bigskip

\noindent
\textbf{Acknowledgements}.
{We thank the anonymous reviewers and PC members of FSTTCS~2021 for their comments.}

\clearpage

{
\bibliographystyle{abbrv}
\bibliography{fgc-jnl-ref}
}
\end{document}